# FEM models and Fourier decomposition of three thick septa

**J. Benesch**

*Thomas Jefferson National Accelerator Facility*
*Newport News, Virginia, USA*
*E-mail*: benesch@jlab.org

ABSTRACT: The CEBAF recirculating linear accelerator uses 26 septa: 8 thin, 17 thick and one Lamberston. We report here on detailed FEM models of one thick current-sheet septum, the Lambertson, and a new thick-septum concept which contains features of both. The last has higher-order field components within a factor of two of the current-sheet septum and much lower current density in the conductor were it used as a "drop-in" replacement in CEBAF. There are accelerator layouts in which the new concept is superior to the current sheet septum in both harmonic content and current density.

# Contents

## 1. Septum magnets

In most accelerators septum magnets are used in injection and extraction [1-9]. CEBAF is a recirculating electron accelerator with two linacs and ten arcs [10]. Electron beams are spread into the required arcs at the end of each linac and coalesced into the following linac at the end of a group of arcs. Of the 26 septa in CEBAF, 12 are used in these spreaders and recombiners, 13 are used in extraction and the last is the Lambertson used to direct beams to three experimental halls. All of these septa except the Lambertson are of the current-sheet type with current densities up to 4 kA/cm2 in the water-cooled copper conductor.

Three magnets were modeled: the Lambertson, a 2 m long thick current-sheet septum used in both the main accelerator and extraction, and a concept developed to reduce the current density needed for a given beam separation and therefore reduce the likelihood of damage due to overheating. CEBAF has experienced one coil fire and has one closed cooling passage in another septa and JLab is finding it difficult to get replacement coils fabricated. The concept is similar to that used in [5,6] in that a steel supplement to a current path is used to reduce the leakage field into the region from which field is to be excluded. It differs from [5,6] in that the current path adjacent to the steel has only one-sixth the amp-turns needed for the deflected beam, with the remainder wound around a more conventional flared pole. The advantage of the concept discussed is that the current density in the conductor is within standard practice and so does not require extraordinary cooling measures.

References [1-9] detail efforts to reduce field in one volume while excluding it from another. They do not address the higher order field content (multipoles) in either region. Opera finite electromagnetic software [11] has a "fit Fourier" command which produces Fourier components of the field. If the mesh is fine enough, a few times that suggested by the sampling theorem, one gets results which are consistent among varying representations (Cartesion, cylindrical, spherical) at the tens of ppm level. In the volume encountered by beam the meshes in the models below are 2 mm quadratic tetrahedra (8 nodes per tetrahedron, one per vertex and one per face.) Evaluation is done on circles of 10 mm diameter along the beam trajectory unless a source (steel or conductor) would be intersected, in which case the radius is reduced and the results scaled by $(r/10)^{-n}$. The circles are at 2.5 mm intervals; the results are summed and divided by 4 to get cm.

## 2. The Lambertson

There are three experimental halls at the west end of the CEBAF complex. Beams may be RF- and septum- separated at passes 1-5 and directed to any one of the halls. A vertical RF separator and two septa allow all three halls to obtain fifth pass beam. This flexibility is allowed by placing the beams at distinct heights as they enter the Lamberston shown in figure 1. Hall A is bent to the right in a passage 20 mm above the Lambertson midplane. Hall B is on the midplane and is not supposed to be deflected or focused in any way. Hall C passage is 20 mm below the midplane and bends to the left. The steel is 2300 mm long.

Issues with the purity of the fields in the Lambertson were discovered when an experiment in Hall B [12] requested a flat beam with horizontal to vertical axis ratio 50:1. The beam ended up tilted rather than horizontal due to the skew quadrupole component in the Lambertson. A detailed finite electromagnetic model of the Lambertson was developed. It was determined that much of the skew quadrupole component in the B line occurred in the vicinity of the A and C coils at the ends of the dipole. Carbon steel tubes were added to the model around the B line to determine how much mitigation could be achieved: about two-thirds. Carbon steel tubes were clamped around the B vacuum vessel at the ends of the magnet. As predicted, skew component dropped but not sufficiently to rotate the flat beam to the horizontal plane. A skew quadrupole was added to the beam line to compensate the remainder.

Since Halls A and C can each receive any one of the five beam momenta available and the magnet is not physically symmetric, twenty-five simulations were run covering all combinations with both halls receiving beam. The ten instances with either A or C off were not simulated as the Experiment Scheduling Committee does its best to avoid this situation. A spreadsheet with all results is available online in Supplemental Material. Table 1 shows a subset of the results.

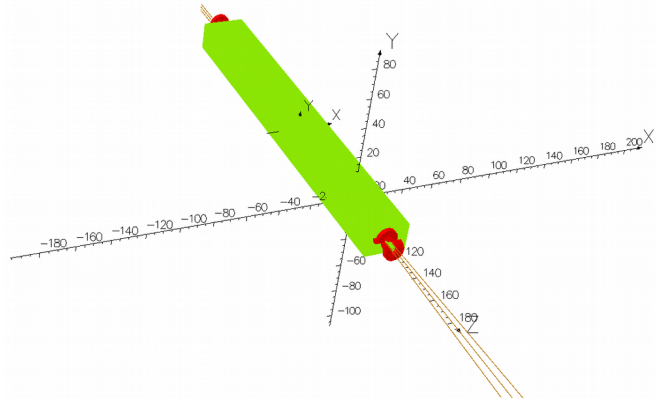

**Figure 1**. Lambertson model. 37M elements, 51M nodes. Three beams stacked vertically at entrance (top). At exit, two beams have been bent horizontally; vertical span retained.

**Table 1**. Fourier components in G-cm for three of 25 Lambertson combinations, normalized to 1 cm radius circles. A(5) denotes fifth pass beam momentum specified desired BdL. C(5) and C(1) analogous. B hole is supposed to be field-free. See Supplemental Material for all 25 cases.

|      | desired BdL | dipole   |      | quadrupole |       | sextupole |       | octupole |       | decapole |      |
|------|-------------|----------|------|------------|-------|-----------|-------|----------|-------|----------|------|
|      |             | Cos0     | Sin0 | Cos1       | Sin1  | Cos2      | Sin2  | Cos3     | Sin3  | Cos4     | Sin4 |
| A(5) | -1020600    | -1022152 | 0    | -468       | -441  | -875      | 107   | -541     | -251  | -531     | -2   |
| B    |             | 0        | 0    | 265        | -1317 | -2        | 3     | -7       | 235   | 1        | 4    |
| C(5) | 1020600     | 1022152  | 0    | -465       | -448  | 869       | -111  | -542     | -243  | 530      | -4   |
| A(5) | -1020600    | -1032840 | 0    | -776       | 244   | -566      | 101   | -551     | -380  | -626     | 0    |
| B    |             | -1056    | 0    | 142        | -635  | 143       | 124   | 11       | 127   | -95      | -14  |
| C(1) | 214500      | 211177   | 0    | 67         | -209  | 234       | -19   | -114     | -16   | 81       | 1    |
| A(1) | -214500     | -211178  | 0    | 66         | -207  | -235      | 18    | -113     | -17   | -82      | -2   |
| B    |             | 1056     | 0    | 141        | -634  | -146      | -121  | 13       | 126   | 93       | 18   |
| C(5) | 1020600     | 1032840  | 0    | -772       | 237   | 559       | -104  | -553     | -372  | 624      | -5   |

The plate through which the B beam passes is 17.8 mm thick. Had these calculations been done before the recent CEBAF energy upgrade it might have been possible to increase this to 20 mm and to make the B beam pipe of carbon steel, increasing shielding of the B aperture and reducing the cross-talk among the three apertures. However, this would have required moving the beam closer to the current sheets in septa upstream, increasing multipoles encountered there. The next section examines one of these septa.

## 3. The YR magnet

The model of the two meter YR current sheet is shown in figure 2. The model has 52M nodes and 37M elements. There are 707M non-zeros in the matrices, a third of Opera's maximum. The magnet steel is top/bottom symmetric so only the upper half need be modeled.

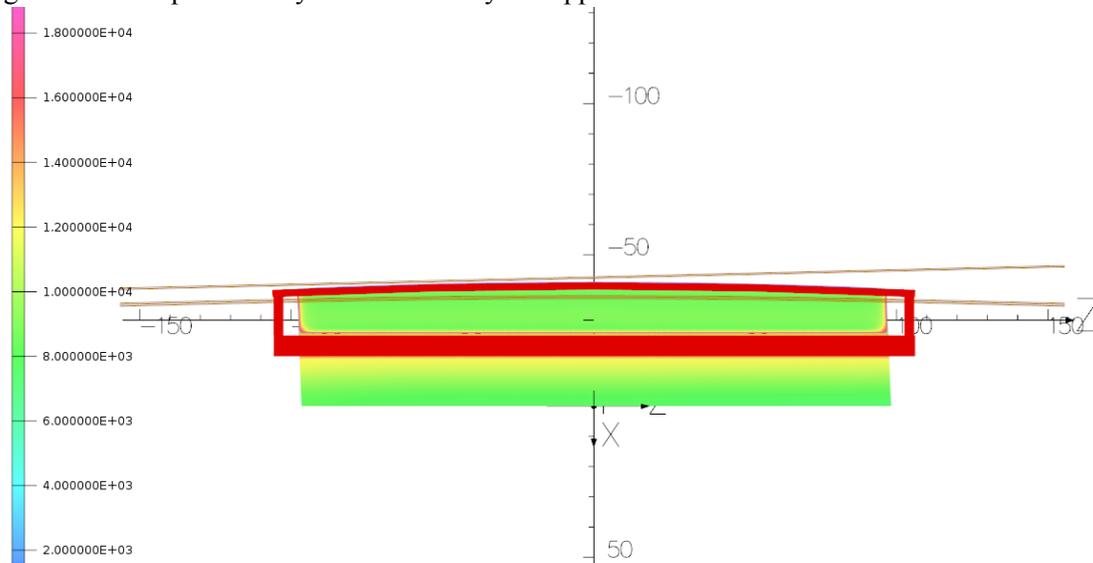

**Figure 2**. YR magnet model. Current path is made of four bars with approximate cross-sections rather than detailing the individual current and water paths on the current sheet, the bar closer to the top of the image. Five orbits are shown, two passing beams with 5 mm spacing and three bent beams at 5 mm intervals. Fourier components are evaluated at 2.5 mm intervals along these.

Fourier components were evaluated on 10 mm, 8 mm and 5 mm radius circles with 60 points on each circle. Circles are at 2.5 mm intervals along the orbits shown. In the table below, black values were evaluated on 10 mm circles and red values on 5 mm circles as larger circles touched the edge of the steel in the model. The coil pack is ~30 mm in width.

**Table 2 Fourier components from YR magnet model evaluated on r=10 mm circles**

| mm to coil | Cos0 | Cos1 | Cos2 | Cos3 | Cos4 |
|---|---|---|---|---|---|
| 15 | -13890 | -10320 | -7160 | -4330 | -2130 |
| 10 | -21000 | -21170 | -17040 | -9870 | -3490 |
| -10 | 1751830 | -15590 | 10230 | -5180 | 1610 |
| -15 | 1746650 | -9680 | 5550 | -2900 | 1310 |
| -20 | 1742820 | -5710 | 2630 | -1170 | 490 |

Since the second and third lines are 10 mm outside and 10 mm inside the 30 mm coil, the beam separation is 50 mm. The harmonic content would be much reduced with 60 mm or 65 mm separation. The dipole term on the passing beam is ~0.01% of that of the deflected beam. The septa in [1-9] attain an order of magnitude improvement on this, 10 ppm of the deflection field at the nominal separation, but harmonic content was not measured. The JLab current septa are being run ~25% above original specification as a consequence of the energy upgrade. The model was also evaluated at 80% and 55% of the case shown above. All harmonics in the bent-beam region were within 0.5% of the current ratio; the span was a bit larger for the passing beams. Steel volume could be reduced without consequence.

**4. The YC septum concept**

The YC concept began as my misunderstanding of a request for a new septum with more robust coil design. The model is shown in figure 3 as simulated utilizing the available symmetries. Four beam paths are shown in figure 4; these were among six along which Fourier components were evaluated. This model is rectangular unlike the YR with its curved current sheet and angled ends. Only two of the beam paths used are physical; the other four are straight so as to provide better comparison with the YR results in Table 2.

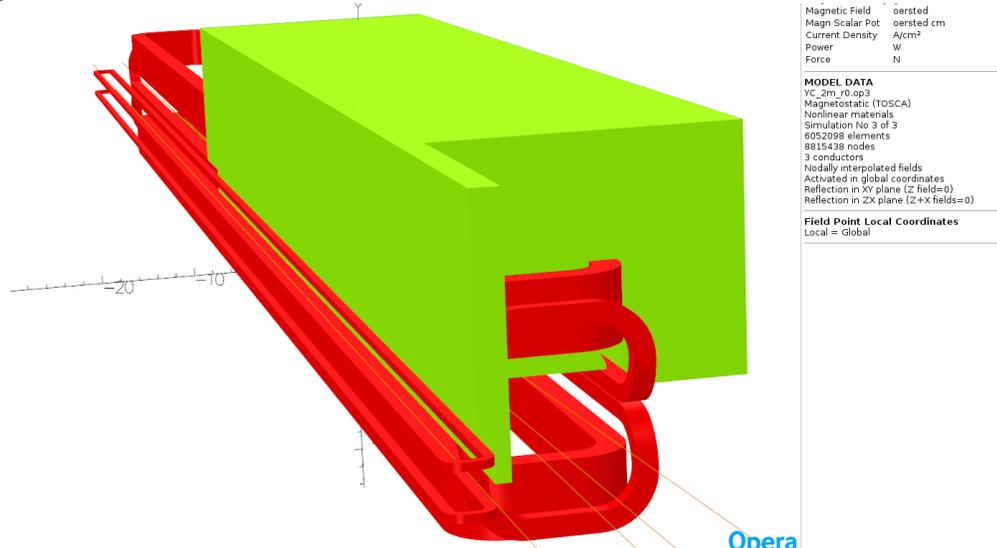

**Figure 3**. YC magnet model. There is a flared pole with a racetrack coil with 24 turns wrapped around it. There is a bedstead coil of 4 turns between the pole faces and adjacent to a 1 cm

steel septum at left. There are a pair of racetrack coils at left to null the residual dipole field outside the steel. The steel septum is 10 mm thick in the passing-beam region and tapers out to 20 mm above these coils. The coils which provide the deflecting field would be fabricated of 6 mm square conductor with 3.5 or 4 mm hole. Current density is 13.9 (16.7) A/mm2 in the model used to generate the field values below, depending on hole size chosen; 338.3 A. Either J is less than half that in the current sheet septum and should allow for four-turn coil units with one cooling water path. The author is not competent to do a detailed thermal analysis. The small outside racetrack has only 15 AT so does not need cooling.

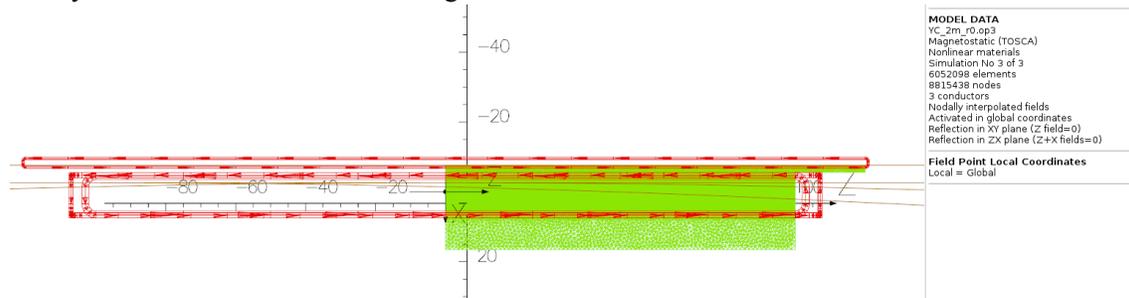

**Figure 4.** The same four trajectories are shown here. Two are straight, at x=-26 and 76 mm, separated by 50 mm. Two are curved, one with the entrance and exit angles equal and placed so orbit closest approach to coil is at -2.6 cm. The second curved orbit has normal incidence at x=-2.6 cm. The two orbits not shown are straight and are at x=-2.1 and -8.1, allowing for 60 mm separation to compare with the +15 mm and -15 mm lines in Table 2.

Finally, in figure 5, |B| is shown on the surface of the full module, truncated at 20000 kG. The flared poles which are sharp in figure 3 are rounded here because the field in the steel is above the plotting cut-off.

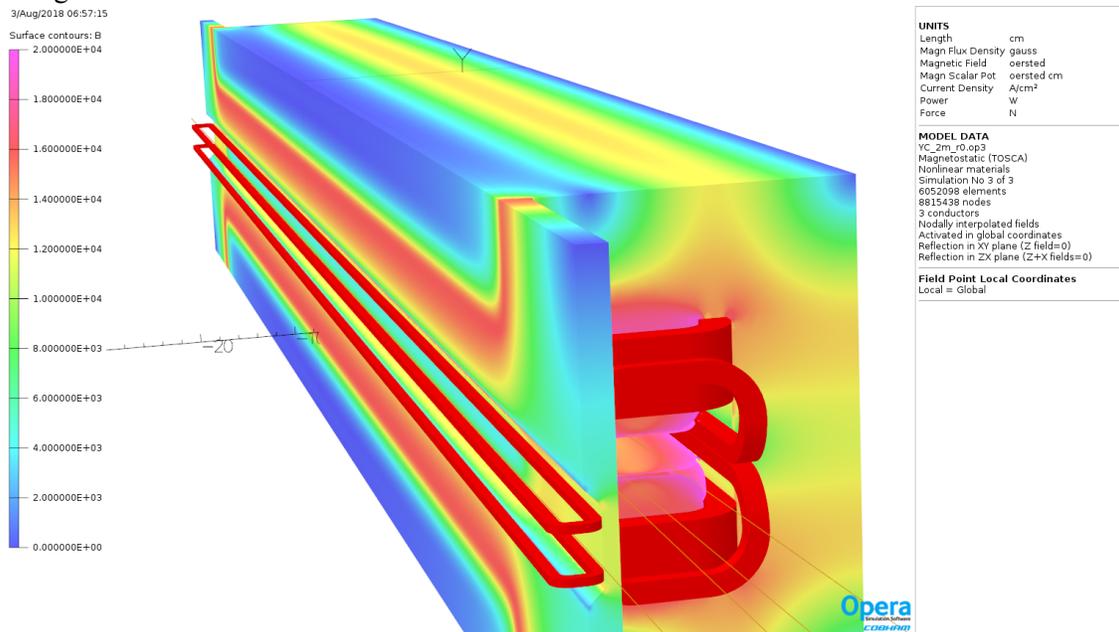

**Figure 5**. YC model with calculated surface field. The taper of the steel septum from 10 mm to 20 mm is more evident here, the yellow region next to the outside coils.

**Table 3. Fourier components from YC magnet model evaluated on r=10 mm circles**

|                       | Cos0     | Cos1   | Cos2  | Cos3   | Cos4  |
|-----------------------|----------|--------|-------|--------|-------|
| entry=exit angle      | -1751537 | -19500 | 17363 | -11553 | 6046  |
| normal entry          | -1756507 | -10349 | 9018  | -5720  | 2984  |
| flat internal x= -2.1 | -1737384 | -13392 | 11718 | -7190  | 3769  |
| flat internal x= -2.6 | -1741799 | -33710 | 30648 | -20710 | 10739 |
| flat external x= -7.6 | 7        | 163    | 274   | -28    | -83   |
| flat external x= -8.1 | 0.36     | -84    | 273   | 35     | 28    |

As in table 2, the red values were evaluated on 5 mm circles as larger radii intercepted the steel, and then scaled to 10 mm radius. The outside racetracks are flat because CEBAF uses rectangular vacuum vessels with septa. If a round beam pipe were used the coils would be modified to reduce the higher Fourier components. If a 40 mm square or round carbon steel vacuum vessel with 2 mm wall were used outside the septum, with a 1 mm air gap, the passing beam would see no significant field. In [6] magnetic stainless steel was used in this way. Using separate power supplies for the racetrack and bedstead coils was explored using Opera's Optimizer module; improvement was under 2% for a 20% increase in current in the bedstead and so not cost effective.

**Table 4. Comparing YR and YC at two beam separations**

|                    | Cos0     | Cos1   | Cos2   | Cos3   | Cos4  |
|--------------------|----------|--------|--------|--------|-------|
| YR 50 mm separation | -21000   | -21170 | -17040 | -9870  | -3490 |
|                    | 1751830  | -15590 | 10230  | -5180  | 1610  |
| YC 50 mm separation | 7        | 163    | 274    | -28    | -83   |
|                    | -1741799 | -33710 | 30648  | -20710 | 10739 |
| YR 60 mm separation | -13890   | -10320 | -7160  | -4330  | -2130 |
|                    | 1746650  | -9680  | 5550   | -2900  | 1310  |
| YC 60 mm separation | 0.36     | -84    | 273    | 35     | 28    |
|                    | -1737384 | -13392 | 11718  | -7190  | 3769  |

Again, values in red were evaluated on 5 mm radius circles and scaled to 10 mm radius. YC model has 15 AT in the external compensation coil.

Finally in table 4 one sees that the YR current sheet septum has lower amplitude multipoles than the YC concept for the bent beam and higher for the passing beam. Referring back to table 3 one sees that the YC concept is superior to the YR if normal entry is possible.

**Conclusion**

Two extant CEBAF septum magnets have been modeled in detail to provide Fourier component input to accelerator modeling with elegant [13]. A concept with thermal advantages but larger harmonic content has been discussed. The concept can provide better magnetic performance than the current sheet if accelerator layout will allow normal beam incidence. The field in the region with circulating beam may be reduced to nil by making the vacuum vessel of magnetic material. The models discussed here and [6] show that this should be standard practice in all "C" septa.

**Acknowledgments**

The YC concept began in my misunderstanding of a discussion with my colleague Mike Tiefenback. This material is based upon work supported by the U.S. Department of Energy, Office of Science, Office of Nuclear Physics under contract DE-AC05-06OR23177.

**Supplemental Material**

Opera opc, res and comi files which will allow those with Opera licenses to duplicate the work presented.

LibreOffice spreadsheets with Fourier components at 2.5 mm intervals which were used to create tables 1-3.